\documentclass[10pt,twocolumn,conference]{IEEEtran}
\usepackage{amsbsy}
\usepackage{times}
\usepackage{amsmath}
\usepackage{latexsym}
\usepackage[dvips ]{graphicx}
\input epsf
\title{\LARGE Joint Model-Order and Step-Size Adaptation using Convex Combinations of Adaptive Reduced-Rank
Filters}
\author{Rodrigo C. de Lamare $\dag$ and V\'itor H. Nascimento \\
Communications Research Group, University of York,
United Kingdom $\dag$ \\ Department of Electronic Systems Engineering, \\
    University of S\~ao Paulo, S\~ ao Paulo, SP, 05508-900 Brazil\\
Emails: rcdl500@ohm.york.ac.uk, vitor@lps.usp.br }

\begin{document}
\maketitle \thispagestyle{empty} \pagestyle{empty}
\begin{abstract}
In this work we propose schemes for joint model-order and
step-size adaptation of reduced-rank adaptive filters. The
proposed schemes employ reduced-rank adaptive filters in parallel
operating with different orders and step sizes, which are
exploited by convex combination strategies. The reduced-rank
adaptive filters used in the proposed schemes are based on a joint
and iterative decimation and interpolation (JIDF) method recently
proposed. The unique feature of the JIDF method is that it can
substantially reduce the number of coefficients for adaptation,
thereby making feasible the use of multiple reduced-rank filters
in parallel. We investigate the performance of the proposed
schemes in an interference suppression application for CDMA
systems. Simulation results show that the proposed schemes can
significantly improve the performance of the existing reduced-rank
adaptive filters based on the JIDF method.   \vspace{-0.35em}

\end{abstract}

\begin{keywords}
{Adaptive filters, convex combinations, model-order selection,
reduced-rank adaptive filters.}
\end{keywords}
\vspace{-0.5em}

\section{Introduction}
\vspace{-0.5em}
In the literature of adaptive filtering algorithms
\cite{haykin}, numerous algorithms with different trade-offs
between performance and complexity have been reported in the last
decades. A designer can choose from the simple and low-complexity
least-mean squares (LMS) algorithms to the fast converging though
complex recursive least squares (RLS) techniques \cite{haykin}. A
great deal of research has been devoted to developing
cost-effective adaptive filters with an attractive trade-off
between performance and complexity and automatic tuning of key
parameters \cite{haykin}. Combination schemes
\cite{jeronimo1}-\cite{vitor} are recent and effective design
approaches, where several filters are mixed in order to obtain an
overall quality improvement. The individual filters are set up so
as to optimize different desirable properties: fast tracking or
low error in steady-state, for example. The combined filter is
able to keep the advantages of all the individual filters,
achieving superior performance despite a higher computational cost
than the single filter approach. The computational complexity
required by combination schemes is due to the use of two or more
adaptive filters in parallel and can become unacceptably high with
large filters \cite{jeronimo1}-\cite{vitor}.

Reduced-rank adaptive filters \cite{strobach}-\cite{delamaretsp09}
are cost-effective techniques when dealing with problems involving
large filters and reduced training. A number of reduced-rank
adaptive filtering methods have been proposed in the last several
years \cite{strobach}-\cite{delamaretsp09}. Among them are
eigen-decomposition-based techniques \cite{strobach}, the multistage
Wiener filter (MSWF) \cite{goldstein}, the auxiliary vector
filtering (AVF) algorithm \cite{avf}, the interpolated reduced-rank
filters \cite{delamarespl1}, the reduced-rank filters based on joint
and iterative optimization (JIO) \cite{delamarespl2} and joint
iterative interpolation, decimation, and filtering (JIDF)
\cite{delamaretsp09}. Key problems with previously reported
reduced-rank adaptive filters are the tuning of the step sizes
and/or the forgetting factors, and the model-order selection
\cite{strobach}-\cite{jidf_echo}.

In this paper, we propose schemes for joint model-order and
step-size adaptation of reduced-rank adaptive filters based on the
JIDF technique \cite{delamaretsp09} which address these drawbacks
of the JIDF scheme. The proposed schemes employ reduced-rank
adaptive filters in parallel operating with different orders and
step sizes, which are exploited by convex combination strategies.
The main reason for choosing the JIDF technique
\cite{delamaretsp09} is that it allows a substantial reduction in
the number of coefficients for adaptation and yields the best
performance among the techniques reported so far. Although the
number of coefficients that are actually adapted is smaller,
leading to better convergence rates and mean-square error, the
complexity of the JIDF is comparable to that of the full-rank LMS.
Other reduced-rank schemes have in general a higher complexity. We
derive least-mean squares (LMS) reduced-rank filters based on the
JIDF method and propose strategies to automatically adjust the
model-order and step sizes used. Without an interpolator, the JIDF
approach may also address the main drawback of combination
schemes, i.e., the increase in the number of elements for
computation, but this possibility will be pursued elsewhere.

This paper is organized as follows. Section II presents the
proposed convex combination schemes of reduced-rank adaptive
filters. Section III is devoted to the derivation of the convex
combiners and Section IV to the derivation of LMS algorithms with
the JIDF approach. Section V presents and discusses the simulation
results and Section VI draws the conclusions of this work.
\vspace{-0.5em}

\section{Proposed Combination Schemes } \vspace{-0.5em}

In this section, we detail the proposed convex combination schemes
of reduced-rank adaptive filters. The basic idea behind these
schemes is to employ a number of parallel sets of transformation
matrices and reduced-rank filters that are jointly optimized and
exploit them via the setting of different model orders and step
sizes. The different model orders and step sizes are determined
{\it a priori}. Since we are interested in this work in adapting
the model order and the step size, then in principle we need at
most $4$ parallel structures for setting upper and lower values
for the model order (or rank) and the step sizes. To this end, we
can build a tree structure where only two structures are combined
at each stage. A block diagram of the first tree structured
scheme, denoted scheme A, is shown in Fig. \ref{fig1}.

Let us now mathematically describe the signal processing performed
by the proposed scheme A. Consider an $M \times 1$ input data
vector ${\boldsymbol r}[i]$ that is processed by reduced-rank
schemes in parallel with different parameters, namely the model
order and the step size. The output of scheme A is given by
\begin{equation}
\begin{split}
y_c[i] & = \lambda_c[i] y_a[i] + (1-\lambda_c[i]) y_b[i] \\
& = \lambda_c[i] \Bigl(\lambda_a[i] y_1[i] + (1-\lambda_a[i]) y_2[i]\Bigr)
\\ & \quad + (1-\lambda_c[i]) \Bigl(\lambda_b[i] y_3[i] + (1-\lambda_b[i])
y_4[i]\Bigr)\\ & = \lambda_c[i] \Bigl(\lambda_a[i] \bar{\boldsymbol
w}_1^H[i] {\boldsymbol S}_{D_1}^H[i] {\boldsymbol r}[i] \\ & \quad
+ (1-\lambda_a[i]) \bar{\boldsymbol w}_2^H[i] {\boldsymbol
S}_{D_2}^H[i] {\boldsymbol r}[i]\Bigr) \\ & \quad + (1-\lambda_c[i])
\Bigl(\lambda_b[i] \bar{\boldsymbol w}_3^H[i] {\boldsymbol
S}_{D_3}^H[i] {\boldsymbol r}[i] \\ & \quad + (1-\lambda_b[i])
\bar{\boldsymbol w}_4^H[i] {\boldsymbol S}_{D_4}^H[i] {\boldsymbol
r}[i]\Bigr) \\ & = {\boldsymbol w}_{\rm eq}^{A,~H}[i] {\boldsymbol
r}[i], \label{output}
\end{split}
\end{equation}
where the equivalent filter ${\boldsymbol w}_{\rm eq}^{A}[i]$ is
given by
\begin{equation}
\begin{split}
{\boldsymbol w}_{\rm eq}^A[i] & = \lambda_c[i] \lambda_a[i]
{\boldsymbol S}_{D_1}[i] \bar{\boldsymbol w}_1[i] \\ & \quad +
\lambda_c[i] (1-\lambda_a[i]) {\boldsymbol S}_{D_2}[i]
\bar{\boldsymbol w}_2[i] \\ & \quad + (1-\lambda_c[i])
\lambda_b[i] {\boldsymbol S}_{D_3}[i] \bar{\boldsymbol w}_3[i] \\
& \quad + (1-\lambda_c[i]) (1-\lambda_b[i]){\boldsymbol
S}_{D_4}[i] \bar{\boldsymbol w}_4[i]
\end{split}
\end{equation}
The main strategy is to set the constituent reduced-rank filters
with an estimate of the lowest and highest ranks $D_{\rm min}$ and
$D_{\rm max}$, respectively, and the smallest and largest step
sizes $\mu_{\rm min}$ and $\mu_{\rm max}$, respectively.
Therefore, the proposed convex combination would be able to
exploit the differences in rank and step size of the constituent
reduced-rank filters and keep their advantages.

Now let us consider a second scheme, denoted scheme B, with the
convex combination of only two structures and shown in Fig.
\ref{fig2}. The output of scheme B is given by
\begin{equation}
\begin{split}
y_c[i] & = \lambda_c[i] y_1[i] + (1-\lambda_c[i]) y_2[i])
\\  & = \lambda_c[i] \bar{\boldsymbol
w}_1^H[i] {\boldsymbol S}_{D_1}^H[i] {\boldsymbol r}[i] +
(1-\lambda_c[i]) \bar{\boldsymbol w}_2^H[i] {\boldsymbol
S}_{D_2}^H[i] {\boldsymbol r}[i] \\ & = {\boldsymbol w}_{\rm
eq}^{B,~ H}[i] {\boldsymbol r}[i], \label{output2}
\end{split}
\end{equation}
where the equivalent filter ${\boldsymbol w}_{\rm eq}^{B}[i]$ is
given by
\begin{equation}
\begin{split}
{\boldsymbol w}_{\rm eq}^{B}[i] & = \lambda_c[i] {\boldsymbol
S}_{D_1}[i] \bar{\boldsymbol w}_1[i] + (1-\lambda_c[i])
{\boldsymbol S}_{D_2}[i] \bar{\boldsymbol w}_2[i]
\end{split}
\end{equation}
The strategy for the scheme B is to set one of the constituent
reduced-rank filters with an estimate of the lowest rank $D_{\rm
min}$ together with the largest step size $\mu_{\rm max}$, whereas
the other uses the highest rank $D_{\rm max}$ and the smallest
step size $\mu_{\rm min}$, respectively. Therefore, the proposed
convex combination would be able to exploit fast adaptation
($D_{\rm min}$ and $\mu_{\rm max}$) with low misadjustment
($D_{\rm max}$ and $\mu_{\rm min}$). \vspace{-0.5em}

\section{The JIDF Reduced-Rank Filter Scheme} \vspace{-0.5em}

We detail the JIDF scheme \cite{delamaretsp09} used as the
constituent of the proposed schemes. Let us now review the JIDF
and its main parameters for the $j$th branch, where $j=1, \ldots,
J$ and $J=2,4$. A block diagram of the JIDF scheme is shown in
Fig. \ref{fig3}, where an interpolator ${\boldsymbol v}_j[i]$ with
$I_j$ coefficients, a decimation unit and a reduced-rank filter
${\boldsymbol w}_j[i]$ with $D_j$ coefficients that are
time-varying are employed. The $M \times 1$ input vector
${\boldsymbol r}[i]$ is filtered by the ${\boldsymbol v}_j[i]$ and
yields the interpolated vector ${\boldsymbol r}_{{\rm I},j}[i]$
with $M$ samples expressed by
\begin{equation}
{\boldsymbol r}_{{\rm I},j}[i] = {\boldsymbol V}^{H}_j[i] {\boldsymbol
r}[i],
\end{equation}
where the $M\times M$ Toeplitz convolution matrix ${\boldsymbol
V}_j[i]$ is given by {
\begin{equation}
{\boldsymbol V}_j[i] =  \left[\begin{array}{c c c c c c c c  }
v_{j,0}^{[i]}  & 0 & \ldots & 0 &  \\
\vdots & v_{j,0}^{[i]} & \ldots & 0 &  \\
v_{j,{I}_{j}-1}^{[i]} & \vdots & \ldots & 0 &  \\
0 & v_{j,{I}_{j}-1}^{[i]}  & \ldots & 0   \\
0  &   0 & \ddots & 0  \\
\vdots &   \vdots & \ddots & \vdots  \\
0 & 0 &  \ldots & v_{j,0}^{[i]} \\
 \end{array}\right]. \nonumber
\end{equation} }


In order to facilitate the description of the scheme, let us
introduce an alternative way of expressing the vector
${\boldsymbol r}_{{\rm I},j}[i]$, that will be useful in the following
through the equivalence:
\begin{equation}
{\boldsymbol r}_{{\rm I},j}[i]={\boldsymbol V}^{H}_j[i]{\boldsymbol
r}[i] = \boldsymbol{ \Re}_{o_{j}}[i]{\boldsymbol v}^{*}_j[i],
\end{equation}
where the $M\times I_j$ matrix $\boldsymbol{ \Re}_{o_{j}}[i]$ with
the samples of ${\boldsymbol r}[i]$ has a Hankel structure
described by
\begin{equation}
\boldsymbol{ \Re}_{o_{j}}[i] = \left[\begin{array}{c c c c c}
r_{0}^{[i]} & r_{1}^{[i]}   & \ldots & r_{{I}_j-1}^{[i]}  \\
r_{1}^{[i]}  & r_{2}^{[i]}   & \ldots & r_{{I}_j}^{[i]}  \\
\vdots & \vdots  & \ddots & \vdots \\
r_{M-1}^{[i]}  & r_{M}^{[i]}  & \ldots & r_{M +{I}_j-2}^{[i]}  \\
 \end{array}\right].
\end{equation}
The dimensionality reduction is performed by a decimation unit
with $D \times M$ decimation matrices ${\boldsymbol D}_{b_{j}}[i]$
that project ${\boldsymbol r}_{{\rm I},j}[i]$ onto $D_j \times 1$
vectors $\bar{\boldsymbol r}_{b_{j}}[i]$ with $b=1, \ldots, B$,
where $D_j=M/L_j$ is the rank and $L_j$ is the decimation factor.
The $D_j \times 1$ vector $\bar{\boldsymbol r}_{b_{j}}[i]$ for
branch $b$ is expressed by
\begin{equation}
\bar{\boldsymbol r}_{b_{j}}[i] = {\boldsymbol D}_{b_{j}}[i]
{{\boldsymbol r}_{{\rm I},j}[i]} = {\boldsymbol D}_{b_{j}}[i]
\boldsymbol{ \Re}_{o_{j}}[i]{\boldsymbol v}^{*}_j[i],
\end{equation}
where the vector $\bar{\boldsymbol r}_{b_{j}}[i]$ for branch $b_j$
is used in the instantaneous minimization of the squared norm of
the error for branch $b_j$
\begin{equation}
e_{b_{j}}[i] = d[i] - \bar{\boldsymbol w}^{H}_j[i]\bar{\boldsymbol
r}_{b_{j}}[i]. \nonumber
\end{equation}
The decimation pattern ${\boldsymbol D}_{b_{j}}[i]$ is selected
according to:
\begin{equation}
{\boldsymbol D}_{b_{{\rm opt}, j}}[i] = {\boldsymbol
D}_{b_{s,j}}[i] ~~ \textrm{when} ~~ b_{s,j} = \arg \min_{1\leq b_j
\leq B} |e_{b_{j}}[i]|^{2},
\end{equation}
where $B$ is the number of decimation branches, which is a
parameter to be set by the designer. We denote
$\bar{\boldsymbol r}_j[i]\leftarrow\bar{\boldsymbol
r}_{b_{\text{opt},j}}  [i]$. After the decimation unit, which
carries out dimensionality reduction, the JIDF scheme employs a
reduced-rank FIR filter $\bar{\boldsymbol w}_j[i]$ with $D_j$
elements to yield the output of the scheme.  A key strategy for
the joint and iterative optimization that follows is to express
the output of the JIDF structure $y_j[i] = \bar{\boldsymbol
w}^{H}_j[i]\bar{\boldsymbol r}_{j}[i]$  as a
function of ${\boldsymbol v}_j[i]$, the decimation matrix
${\boldsymbol D}_{b_j}[i]$ and $\bar{\boldsymbol w}_j[i]$ as
follows:
\begin{equation}
\begin{split}
y_j[i] & = \bar{\boldsymbol w}^{H}_j[i] {\boldsymbol
S}_{D_j}^H[i]{\boldsymbol r}[i] = \bar{\boldsymbol w}^{H}_j[i]
{\boldsymbol D}_{b_{j}}[i]\boldsymbol{\Re}_{o_{j}}[i] {\boldsymbol
v}^{*}_j[i] = {\boldsymbol v}^{H}_j[i]{\boldsymbol u}_j[i]
\end{split},
\end{equation}
where ${\boldsymbol u}_j[i]=
\boldsymbol{\Re}_{o_j}^T[i]{\boldsymbol D}^T_{b_j}[i]
\bar{\boldsymbol w}^{*}_j[i]$ is an ${I}_j\times 1$ vector. The
expression in (10) indicates that the dimensionality reduction
carried out by the proposed scheme depends on finding appropriate
${\boldsymbol v}_j[i]$, ${\boldsymbol D}_{b_j}[i]$ for
constructing ${\boldsymbol S}_{D_j}[i]$. In the next section, we
will develop adaptive algorithms for adjusting the coefficients of
${\boldsymbol v}_j[i]$ and $\bar{\boldsymbol w}_j[i]$ for
determining the best ${\boldsymbol D}_{b_j}[i]$ iteratively.
\vspace{-0.5em}

\section{Proposed Adaptive Algorithms} \vspace{-0.5em}

In this section, we develop adaptive LMS algorithms for the
proposed convex combination scheme. The key feature of the
proposed algorithms is the joint and iterative optimization of the
filters, the decimation unit and the convex combiners.

The algorithms can be derived by minimizing the cost function
\begin{equation}
\begin{split}
{\mathcal C}({\boldsymbol v}_j[i], D_{b_j}[i],\bar{\boldsymbol
w}_j[i], \lambda_u[i]) = E[| d[i] - y_c[i]|^2, \label{cost}
\end{split}
\end{equation}
where ${\boldsymbol v}_j[i]$, $D_{b_j}[i]$ and $\bar{\boldsymbol
w}_j[i]$ are the interpolators, the decimators and the
reduced-rank filters of the $j$th constituent filtering scheme,
and $\lambda_u[i]$ where $u=a, b, c$ are the generic combiners.
The mixing parameters $\lambda_u[i]$ in the tree structure of the
schemes depicted in Figs. \ref{fig1} and \ref{fig2} are updated
via auxiliary variables $u[i]$ ($u[i]=a[i], b[i], c[i]$) and a
sigmoid function, as in $\lambda_u[i] = \frac{1}{1+e^{-u[i]}}$.

We derive next the expressions for scheme A. The expressions for
scheme B can be derived in a similar way. Substituting the output
of scheme A \eqref{output} into the cost function, minimizing the
cost function with respect to $D_{b_j}[i]$ and computing the
instantaneous gradients of the cost function with respect to
${\boldsymbol v}_j[i]$, $\bar{\boldsymbol w}_j[i]$, we get the
following JIDF recursions for $j=1, \ldots, J$
\begin{equation}
b_{{\rm opt},j} = \arg \min_{1 \leq b \leq B} |e_{b_j}[i]|^2,
\end{equation}
where the error signal used in the decimation unit is $e_{b_j}[i]
= d[i] - \bar{\boldsymbol w}_j^H[i] {\boldsymbol D}_{b_j}[i]
{\boldsymbol \Re}_{o_{j}}[i] {\boldsymbol v}_j^*[i]$. After the
selection of the best branch, the error signal becomes $e_j[i]
\leftarrow  e_{b_{{\rm opt},j}}[i]$. The recursions for
${\boldsymbol v}_j[i]$ and $\bar{\boldsymbol w}_j[i]$ are
\begin{equation}
{\boldsymbol v}_j[i+1] = {\boldsymbol v}_j[i] + \eta_j e_j^*[i]
{\boldsymbol u}_j[i],
\end{equation}
\begin{equation}
{\boldsymbol w}_j[i+1] = {\boldsymbol w}_j[i] + \mu_j e_j^*[i]
{\boldsymbol r}_j[i],
\end{equation}
where the $I_j \times 1$ vector ${\boldsymbol u}_j[i] =
{\boldsymbol \Re}_{o_j}^T[i] {\boldsymbol D}_{b_{{\rm opt}, j}}^T[i]
\bar{\boldsymbol w}_j[i]$ is the regressor for the update
recursion of the interpolator ${\boldsymbol v}_j[i]$ and the $D_j
\times 1$ vector ${\boldsymbol r}_j = {\boldsymbol D}_{b_{{\rm
opt}, j}}[i] {\boldsymbol \Re}_{o_j}[i] {\boldsymbol v}_j^*[i]$ is
the regressor for the update equation of $\bar{\boldsymbol
w}_j[i]$ and the error signal is $e_j[i] = d[i] - \bar{\boldsymbol
w}_j^H[i] {\boldsymbol D}_{b_{{\rm opt}, j}}[i] {\boldsymbol
\Re}_{o_j}[i] {\boldsymbol v}_j^*[i]$.

Now, we need to derive the recursions for the convex combiners
$\lambda_u[i]$ for $u=a,b,c$. Computing the gradient of the cost
function in (\ref{cost}) with respect to $u[i]$ for $u=a$ we
obtain the following recursion
\begin{equation}
\begin{split}
a[i+1] & = a[i] - \mu_a \frac{\partial {\mathcal C}({\boldsymbol
v}_j[i], D_{b_j}[i],\bar{\boldsymbol w}_j[i],
\lambda_u[i])}{\partial a[i]} \\
& = a[i] - \mu_a \frac{\partial {\mathcal C}({\boldsymbol v}_j[i],
D_{b_j}[i],\bar{\boldsymbol w}_j[i], \lambda_u[i])}{\partial
\lambda_a[i]} \frac{\partial \lambda_a[i]}{\partial a[i]} \\
& = a[i] + \mu_a ( y_1[i] - y_2[i])^* \lambda_a[i] (1-
\lambda_a[i]) e_a[i],
\end{split}
\end{equation}
where the error signal for this combiner is $e_a[i] = d[i] -
y_a[i]$ and the combiner is $\lambda_a[i] =
\frac{1}{1-e^{-a[i]}}$. Following this approach, we can obtain the combiner for $u=b$:
\begin{equation}
\begin{split}
b[i+1] & = b[i] + \mu_b  ( y_3[i] - y_4[i])^* \lambda_b[i] (1-
\lambda_b[i]) e_b[i],
\end{split}
\end{equation}
where the error signal for this combiner is $e_b[i] = d[i] -
y_b[i]$ and the combiner is $\lambda_b[i] =
\frac{1}{1-e^{-b[i]}}$. The recursion for the last combiner in the
tree structure is
\begin{equation}
\begin{split}
c[i+1] & = c[i] + \mu_c ( y_a[i] - y_b[i])^* \lambda_c[i] (1-
\lambda_c[i]) e_c[i],
\end{split}
\end{equation}
where the error signal for this combiner is $e_c[i] = d[i] -
y_c[i]$ and the combiner is $\lambda_c[i] =
\frac{1}{1-e^{-c[i]}}$. The complexity of the existing algorithms
is $2M$ additions and $2M+1$ multiplications for the full-rank LMS
algorithm, and $4M+5$ additions and $4M+6$ multiplications for the
full-rank LMS algorithms with convex combination (full-rank-CLMS).
For the JIDF scheme $M(I-1)+(B+1)D+2I$ additions and $MI+(B+2)D$
multiplications are required, where $I$ and $D$ are the lengths of
the interpolator and the reduced-rank filter. The complexity of
the proposed schemes A and B with the JIDF is
$\sum_{j=1}^{J}\bigl[M(I_j-1)+(B+1)D_j+2I_j\bigr]$ additions and
$\sum_{j=1}^{J}\bigl[MI_j +(B+2)D_j\bigr]$ multiplications. We can
reduce the complexity of the JIDF by setting small values for $I$,
$D$ and $B$. However, the key advantage is a substantial reduction
in the number of coefficients that need to be estimated, from $M$
to $I+D$, where $I+D \ll M$.  This allows a much faster adaptation
rate and smaller excess mean-square error, compared to full-rank schemes. \vspace{-0.5em}

\section{Simulations} \vspace{-0.5em}

We assess the performance of the proposed and existing schemes for
interference suppression in CDMA systems. We compare the full-rank
scheme with the convex combination scheme of \cite{jeronimo1}, the
JIDF \cite{delamaretsp09}, the JIDF with the proposed schemes A and
B, and the optimal linear minimum mean-squared error (MMSE) filter
that is computed with the knowledge of the channels, the signature
sequences of all users, and the noise variance at the receiver. All
techniques are equipped with LMS algorithms. Consider the downlink
of a synchronous DS-CDMA system with $K$ users, $N$ chips per symbol
and $L_{p}$ paths. Assuming that the channel is constant during each
symbol interval, the received signal after filtering by a chip-pulse
matched filter and sampled at the chip rate yields the $M\times 1$
received vector
\begin{equation}
{\bf r}[i] = \sum_{k=1}^{K}A_{k}b_{k}[i]{\bf C}_{k} {\bf h}_{k}[i]
 + \boldsymbol{\eta}_{k}[i] + {\bf n}[i],
\end{equation}
where $M=N+L_{p}-1$, ${\bf n}[i] = [n_{1}[i]
~\ldots~n_{M}[i]]^{T}$ is the complex Gaussian noise vector with
$E[{\bf n}[i]{\bf n}^{H}[i]] = \sigma^{2}{\bf I}$, where
$(\cdot)^{T}$ and $(\cdot)^{H}$ denotes transpose and Hermitian
transpose, respectively. The operator $E[\cdot]$ stands for
ensemble average, $b_{k}[i] \in \{\pm1 j\pm1\}/\sqrt{2}$ is the
symbol for user $k$ with $j^{2}=-1$, $\boldsymbol{\eta}[i]$
represents the ISI, the amplitude of user $k$ is $A_{k}$, the
channel vector is ${\bf h}[i] = [h_{0}[i]~ \ldots~
h_{L_{p}-1}[i]]^{T}$ and the $M \times L_{p}$ convolution matrix
${\bf C}_{k}$ contains one-chip shifted versions of the signature
sequence for user $k$ given by ${\bf s}_{k} = [a_{k}(1) \ldots
a_{k}(N)]^{T}$. The linear receiver observes $M=N+L_p-1$ samples
per symbol and employs one of the analyzed and proposed schemes,
which provides an estimate of the desired symbol as given by $
\hat{b}_1[i] =  \textrm{sgn}\big(\Re\big[{ y}_{c}[i]\big]\big) +
 j \textrm{sgn}\big(\Im\big[{ y}_{c}[i]\big]\big)$,
where $\Re(\cdot)$ and $\Im(\cdot)$ select the real and imaginary
parts, respectively, $\textrm{sgn}(\cdot)$ is the signum function,
and we consider user $1$ as the desired one. In our simulations,
we use $L_p=9$ and $N=32$, the channels are time-varying with
complex gains computed with Clarke's model \cite{rappa}, and have
$3$ paths with profile $0$, $-3$ and $-9$ dB with spacing between
paths randomly distributed between $0$ and $2$ chips.

\begin{figure}[!htb]
\begin{center}
\def\epsfsize#1#2{0.95\columnwidth}
\epsfbox{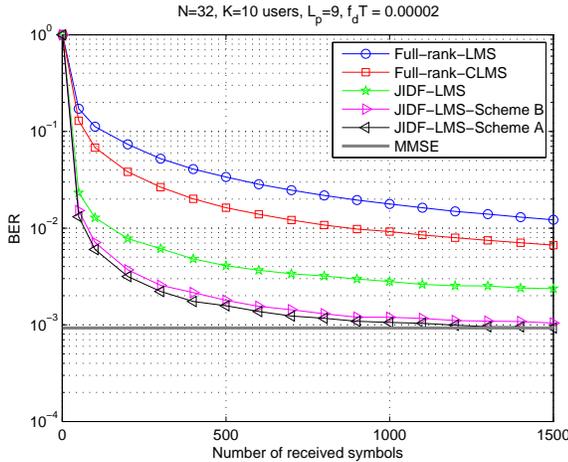} \vspace{-1.5em}\caption{\footnotesize BER
performance versus number of received symbols. Parameters of LMS
algorithms: $\mu=0.05$ (full-rank), $\mu_1=0.01$, $\mu_2=0.25$ and
$\mu_a=0.25$ (combination of full-rank), $B=8$, $I=3$, $D=4$,
$\eta=0.005$ and $\mu=0.01$ (JIDF), $B=8$, $D_1=3$, $I_1=3$,
$\eta_1=0.01$ and $\mu_1=0.1$, $D_2=6$, $I_2=6$, $\eta_2=0.01$ and
$\mu_2=0.1$, $D_3=3$, $I_3=3$, $\eta_3=0.0075$, $\mu_3=0.01$,
$D_4=6$, $I_4=6$, $\eta_4=0.0075$, $\mu_4=0.01$ (JIDF with Scheme
A), $B=8$, $D_1=3$, $I_1=3$, $\eta_1=0.01$ and $\mu_1=0.1$, $D_2=6$,
$I_2=6$, $\eta_2=0.0075$ and $\mu_2=0.01$ (JIDF with Scheme B).}
\vspace{-1.5em}\label{fig4}
\end{center}
\end{figure}

In the first experiment, we assess the bit error ratio (BER)
performance against the received symbols. Packets of $1500$ QPSK
symbols are transmitted and the curves are averaged over $100$
runs. The results depicted in Fig. \ref{fig4} show that the
proposed schemes A and B with the JIDF scheme obtain significantly
better performance than the JIDF without any combination. The use
of the proposed schemes is able to jointly adjust the best model
order and exploit the different step sizes for adaptation. In
terms of computational complexity, the scheme B is more attractive
as it has a performance very close to that of scheme A, but
employs only $2$ JIDF constituent structures as opposed to scheme
A, which uses a combination of $4$ filters.

\begin{figure}[!htb]
\begin{center}
\def\epsfsize#1#2{0.95\columnwidth}
\epsfbox{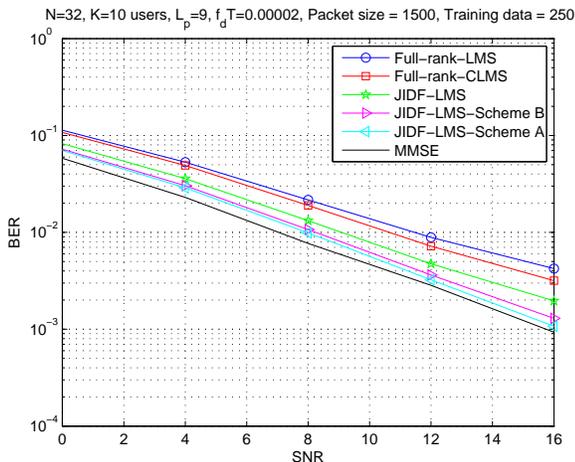} \vspace{-1.5em}\caption{\footnotesize BER
performance versus SNR. Parameters are optimized.}
\vspace{-1.5em}\label{fig5}
\end{center}\vspace{-1.0em}
\end{figure}

In the second experiment, we assess the BER performance against
the signal-to-noise-ratio (SNR) defined as $E_b/N_0$. Packets of
$1500$ QPSK symbols are again transmitted and the curves are
averaged over $100$ runs. The results depicted in Fig. \ref{fig5}
show that the proposed schemes A and B with the JIDF scheme can
obtain significantly better performance than the JIDF without any
combination for different values of SNR. From the results, we
conclude that it might be more attractive to use Scheme B as it
achieves a performance very close to Scheme A with half the
complexity.

\section{Conclusions}
\vspace{-0.2em}
 We proposed convex combination schemes for joint
model-order and step-size adaptation of reduced-rank adaptive
filters based on the JIDF method. The proposed schemes employ
reduced-rank adaptive filters in parallel operating with different
orders and step sizes, which are exploited by convex combination
strategies. We investigated the performance of the proposed
schemes in an interference suppression application for CDMA
systems. Simulations showed that the proposed schemes
significantly improve the performance of the existing reduced-rank
adaptive filters based on the JIDF method. 

\end{document}